\documentclass{jfm}
\usepackage{amsmath,amsfonts}
\usepackage{subcaption}
\usepackage{natbib}
\usepackage{color}

\DeclareMathOperator{\sech}{sech}

\shorttitle{The viscous Holmboe instability}
\shortauthor{J. P. Parker, C. P. Caulfield and R. R. Kerswell}

\title{The viscous Holmboe  instability for smooth shear and density profiles}

\author{J. P. Parker\aff{1}\corresp{\email{jeremy.parker@damtp.cam.ac.uk}}, C. P. Caulfield\aff{2,1} \and R. R. Kerswell\aff{1}}

\affiliation{\aff{1}Department of Applied Mathematics and Theoretical Physics, University of Cambridge,
Wilberforce Rd, Cambridge CB3 0WA, UK
\aff{2} BP Institute for Multiphase Flow, University of Cambridge, Madingley Rise, Madingley Road, Cambridge CB3 0EZ, UK}

\begin{document}
\maketitle

\begin{abstract}
    The Holmboe wave instability is one of the classic examples of a stratified shear instability,
    usually explained as the result of a resonance between a gravity wave and a vorticity wave.
    Historically, it has been studied by linear stability analyses at infinite Reynolds number, $Re$,
    and by direct numerical simulations at relatively low $Re$ in the regions known to be unstable
    from the inviscid linear stability results. In this paper, we perform linear stability analyses of the
    classical `Hazel model' of a stratified shear layer (where the background velocity and density distributions are assumed to take the functional form of hyperbolic tangents with different characteristic vertical scales) over a range
    of different parameters at finite $Re$, finding new unstable regions of parameter space.
    In particular, we find instability when the Richardson number is everywhere greater than $1/4$,
    where the flow would be stable
    at infinite $Re$ by the Miles-Howard theorem. We find unstable modes with no critical layer,
    and show that despite the necessity of viscosity for the new instability, the growth rate relative
    to diffusion of the background profile is maximised at large $Re$.
    We use these results to shed new light on the
    wave-resonance and over-reflection interpretations of stratified shear instability.
    We argue for a definition of Holmboe  instability as being characterised by propagating vortices
    above or below the shear layer, as opposed to any reference to sharp density interfaces.
\end{abstract}

\section{Introduction}
Stably stratified shear flows are ubiquitous in the oceans and atmosphere. Their instabilities
are believed to be relevant to a variety of geophysical processes, and understanding them is important,
for example, in the irreversible
mixing of fluid of different densities in the abyssal ocean to close ocean energy budgets.
The classical example of a shear instability is the Kelvin-Helmholtz instability (KHI) of a uniform sheet
of vorticity. Generally, this instability is damped when a stable stratification is introduced,
and the linear instability is no longer found when the minimum Richardson number, quantifying the strength
of stratification to shear effects, exceeds one quarter \citep{drazin_1958,miles_stability_1961,howard_note_1961}.
However, if a sharp density interface is considered, a qualitatively different, propagating instability
is instead found \citep{holmboe_1962,hazel1972numerical}. This so-called Holmboe wave instability (HWI),
or just `Holmboe instability', is believed to be due to an interaction between internal gravity waves
on the density interface and vorticity waves on either side of the shear layer.
It is hypothesized to be important for ocean mixing \citep{salehipour_caulfield_peltier_2016}, as sharp interfaces
are naturally occurring at high Prandtl numbers.

One important result is the Miles-Howard theorem \citep{miles_stability_1961,howard_note_1961},
which states that in the inviscid case, a stratified shear profile is linearly stable so long as
the local or gradient Richardson number $Ri_g$ (defined precisely below in  section \ref{sec:equations}) is everywhere greater than one quarter. For flows in which HWI is usually
studied, including the piecewise linear profile of \citet{holmboe_1962} and the one-sided profile of \citet{baines1994mechanism},
as well as the smooth profile studied by \citet{hazel1972numerical}, $Ri_g$ is
vanishingly small away from the shear layer, so the theorem does not apply, despite arbitrarily large \emph{bulk} Richardson numbers (also defined more precisely in section \ref{sec:equations}).
On the other hand, when the bulk Richardson number $Ri_b$ is small, the internal waves are not strong, and
so KHI is preferred over HWI.

Though the Miles-Howard theorem is only proven for inviscid flows, a Richardson number of one quarter is
often employed as a criterion for stability in oceanography and related fields.
It is argued from this that a density interface must be narrow compared with the shear layer for HWI
to be present \citep{thorpe_2005}, quantified by the ratio $R$ of shear layer thickness to buoyancy interface thickness being high.
However, \citet{miller_lindzen_1988}
showed that it is possible to have shear instabilities when $Ri_g>1/4$ everywhere if viscosity is introduced.
This leads to the possibility that HWI exists even when $R$ is of order one,
when $Ri_g>1/4$, at finite Reynolds number.
Such an instability was demonstrated, for a single specific choice of parameters, by the authors previously in \citet{pck19}.
This could have profound implications for our understanding of geophysical processes,
since HWI is known to have very different mixing properties to KHI \citep{salehipour_caulfield_peltier_2016}.

In addition to a succinct proof of the Miles-Howard theorem, \citet{howard_note_1961} also
proves an important result, now called the Howard semicircle theorem. This states that for an
inviscid, parallel, stratified shear flow, the complex phase speed of any unstable mode
must be located in a semicircle centred about the median velocity on the real axis, with radius
of half the velocity difference. Though difficult to interpret directly, this has the immediate corollary
that the phase speed of any instability must lie between the maximum and minimum velocities of the flow.
For a smooth velocity profile, this means that there certainly exists a critical layer,
at which the phase velocity equals the flow velocity, and the Taylor-Goldstein equation (see section \ref{sec:equations})
becomes singular. The behaviour of instabilities at the critical layer is a well studied field \citep{maslowe1986, troitskaya_1991,churilov_shukhman_1996},
and leads to the over-reflection hypothesis discussed below. However, the semicircle theorem
is again only proven for inviscid flows, and we shall see that it does not generalise when viscosity is
taken into account.

Two different physical interpretations of stratified shear instabilities exist in the literature.
The first, suggested originally by \citet{taylor1931}, developed by
\citet{garcia_1956,cairns_1979,caulfield_1994} and \citet{baines1994mechanism}, and reviewed in detail by \citet{carpenter_2013},
is the idea that a pair of waves can phase-lock and mutually amplify one another if configured correctly.
This leads to the classification of three canonical instabilities: KHI, the resonance of two vorticity waves;
HWI, the resonance of a vorticity and an internal wave; and the so-called `Taylor-Caulfield' instability \citep{taylor1931, caulfield1995experimental}, the resonance of
two internal waves. In practice, the distinction between these is not clear cut \citep{carpenter_2010, eaves_balmforth_2019}.
In this paper, we shall argue that any instability with zero phase speed in flows with a single density interface
should be defined as KHI and any instability with a propagating localised vortex should be defined
as HWI. The reason for this proposed classification is based on the qualitative nonlinear evolutions, as will become clear in section \ref{sec:nonlinear}.

There is good evidence that an interaction of a gravity and a vorticity wave is responsible for (at least inviscid) HWI.
For instance, \citet{alexakis2005holmboe} discovered additional bands of instability at higher $Ri_b$,
which seem to be caused by the resonance of a higher order gravity wave harmonic with the vorticity wave.
In the piecewise linear model, directly considering the interaction of the two waves in isolation
leads to accurate prediction of the band of instability \citep{caulfield_1994, baines1994mechanism}.
One major problem with this wave-resonance description is that it does not account for the Miles-Howard theorem.
It is not clear why, with a broader density interface, the waves should no longer be able
to resonate and cause instability. Further, though KHI seems to be related to an interaction
of two vorticity waves, the theory has not yet been able to explain the damping of this instability
as Richardson number is increased.

A different perspective, developed by R. S. Lindzen and coauthors, and reviewed in \citet{lindzen1988instability},
is based on the idea that when the local Richardson number is less than one quarter, the critical layer
of a normal-mode wave incident on a stratified shear layer will `over-reflect', and in the correct configuration,
this may lead to exponential growth. This theory, though harder to understand intuitively than
the wave-resonance picture, is attractive as it explicitly includes the Miles-Howard criterion. However,
\citet{smyth1989transition} showed that while wave over-reflection could accurately predict KHI
and HWI in isolation, near the transition between the two, the theory was insufficient.
In particular, there exist regions of parameter space where KHI exists, so the critical layer is
located where the velocity vanishes, and yet $Ri_g>1/4$ here so over-reflection is not expected.

In this paper, we perform linear stability analyses over a wide range of parameters for the `Hazel model' \citep{hazel1972numerical}, including viscosity which has usually been omitted in the past \citep{hazel1972numerical,smyth1989transition,alexakis2005holmboe,alexakis2009stratified}.
As well as finding a clear continuation of the classic inviscid HWI at values of $R$ for which it is known to exist, we also find instability at much lower $R$, with growth rates which vanish
as Reynolds number is increased. We term this new instability the viscous Holmboe instability (VHI), and demonstrate that it has many similarities to the classic HWI.
Our results suggest that while the wave interaction theory gives a useful interpretation of the phenomenology,
neither this nor the over-reflection theory is useful as a necessary or sufficient criterion to predict instability.
We shall see that results from inviscid theory are not only not strictly valid, but are not relevant in
the viscous case, even in situations where the Reynolds number is sufficiently high that a `frozen flow'
approximation is valid, and so the diffusion of the background velocity and density distributions is not thought to be significant.

The remainder of the paper is organised as follows. In section \ref{sec:equations}, we present
the assumptions made and the equations solved. In section \ref{sec:results}, linear stability
analyses are presented over a wide range of different parameters, and the fastest growing
Holmboe modes are tracked and discussed as Reynolds number and $R$ are varied.
Section \ref{sec:nonlinear} shows the nonlinear evolution of VHI
at parameter values for which we expect it to grow fastest, and we compare this against the evolution
of the classic, inherently inviscid HWI.
In section \ref{sec:conclusion},
the results are discussed with particular emphasis on interpretation through wave-resonance and over-reflection.

\section{Equations}
\label{sec:equations}
In this paper, we consider only two-dimensional perturbations to the background flow.
This is a common assumption, by appealing to the results of \citet{squire_1933} and \citet{yih_1955}, who
showed that any three-dimensional normal mode can be associated with a two-dimensional one with
smaller Richardson number and larger Reynolds number. However, this is not necessarily sufficient
to show that the fastest growing mode is always a two-dimensional one \citep{smyth_peltier_1990}.
We discuss this further in section \ref{sec:conclusion}.

An infinitesimal normal-mode perturbation with vertical velocity $w(x,z,t) = \hat{w}(z)e^{i k(x-ct)}$ to
an inviscid Boussinesq flow with velocity profile $U(z)$ and buoyancy profile $B(z)$
must satisfy the well-known Taylor-Goldstein equation
\begin{equation}
    (U-c)\left(\partial_z^2-k^2\right)\hat{w}-U_{zz}\hat{w}=-\frac{B_z}{U-c}\hat{w}.
    \label{eq:TG}
\end{equation}
Here $k$ is the streamwise wavenumber of the perturbation, and $c=c_r+ic_i$ is the complex phase speed, so that
the growth rate of a disturbance is given by $\sigma=kc_i$.

When kinematic viscosity $\nu^*$ and diffusivity of the buoyancy distribution $\kappa^*$ are taken into account, (\ref{eq:TG}) becomes the more complicated pair of equations
\begin{align}
    \begin{split}
        (U-c)\left(\partial_z^2-k^2\right)\hat{w}-U_{zz}\hat{w}&=ik\hat{b}+\frac{1}{ik}\frac{1}{Re}\left(\partial_z^2-k^2\right)^2 \hat{w},\\
        (U-c)\hat{b}+\frac{1}{ik}B_z\hat{w}&=\frac{1}{ik}\frac{1}{Pr Re} \left(\partial_z^2-k^2\right)\hat{b},
        \quad Re\equiv \frac{U^*_0 d^*_0}{\nu^*}; \ Pr \equiv \frac{\nu^*}{\kappa^*},
        \label{eq:VTG}
    \end{split}
\end{align}
where length scales and time scales have been non-dimensionalised using the 
half-depth $d^*_0$ of the shear layer, and half the velocity difference  $U^*_0$ across the shear layer, leading to a conventional
definition of the Reynolds number $Re$, and 
$Pr$
is the
usual Prandtl number.

Following \citet{hazel1972numerical} and many subsequent authors, we consider the `Hazel' model for the background velocity and buoyancy  distributions:
\begin{equation}
U(z)=\tanh z, \ B(z) = \frac{J}{R} \tanh Rz;
\ R\equiv \frac{d^*_0}{\delta^*_0};
\ J\equiv \frac{B_0^*  d_0^*}{U_0^{*2}}, \label{eq:hazel}
\end{equation}
where $\delta^*_0$ is the (dimensional) half-depth of the background
buoyancy layer with half difference  $B_0^*$ and $J$ is the  bulk Richardson number.
This is an extension of the Holmboe model \citep{holmboe_1960},
which has $R=1$ and is attractive because the stability boundary can be found analytically \citep{miles_stability_1961}.
It is  close to the self-similar error function profile expected for a diffusing stratified shear layer when $Pr=R^2$
\citep{smyth_klaasen_peltier}.
It is important to note that these profiles are not steady solutions of the viscous Boussinesq equations,
but we make the `frozen flow' approximation \citep{smythcarpenter2019} which is valid when $\sigma\gg\frac{1}{Re}$.
This inequality is not always satisfied by the instabilities we find, as discussed in section \ref{sec:re}.

The gradient Richardson number $Ri_g$, defined as
\begin{equation}
    Ri_g(z) \equiv \frac{dB/dz}{\left(dU/dz\right)^2} = J \frac{\sech^2 Rz}{\sech^4 z},
\end{equation}
for the Hazel model flow
which means (for this particular flow) that at the centreline, $Ri_g(0) = J$.
For $R\leq\sqrt{2}$, $J$ is the minimum of $Ri_g$, for $\sqrt{2}<R<2$, there are two minima of $Ri_g<J$
either side of local maximum $z=0$, and for $R\geq 2$, $Ri_g\to0$ as $z\to\infty$ and $z=0$ is a global maximum \citep{alexakis2005holmboe}.
From the Miles-Howard theorem, we then deduce that inviscid HWI at arbitrarily large $J$ is only possible when $R\geq 2$.
In fact, \citet{alexakis2007} showed that HWI only exists at all for $R\geq 2$, despite the possibility of instability
at $J>1/4$ when $\sqrt{2}<R<2$.

The solution of (\ref{eq:VTG}) is performed using a MATLAB code from \citet{smythcarpenter2019}.
The method is to construct a large matrix eigenvalue problem, using evenly spaced finite differences.
This is a mature code, and additionally the existence of viscous Holmboe was confirmed in DNS of the Boussinesq
equations at finite $Re$ and $Pr=1$ \citep{pck19}.
The boundary conditions are that $\frac{\partial\hat{w}}{\partial z}=\frac{\partial\hat{b}}{\partial z}=0$,
i.e. frictionless, insulating boundaries,
at $z=\pm L_z$, though all of the instabilities we discuss here are centred around the shear layer, and changing
the boundary conditions would not qualitatively affect the results.
All linear stability results are found using $768$ finite difference points in the vertical direction,
except for the $L_z=20$ case which used $1024$ points,
and the figures are generated from a $48\times48$ grid of calculated growth rates.

\section{Linear stability analyses}
\label{sec:results}
\begin{figure}
    \centering
    \includegraphics[width=0.8\textwidth]{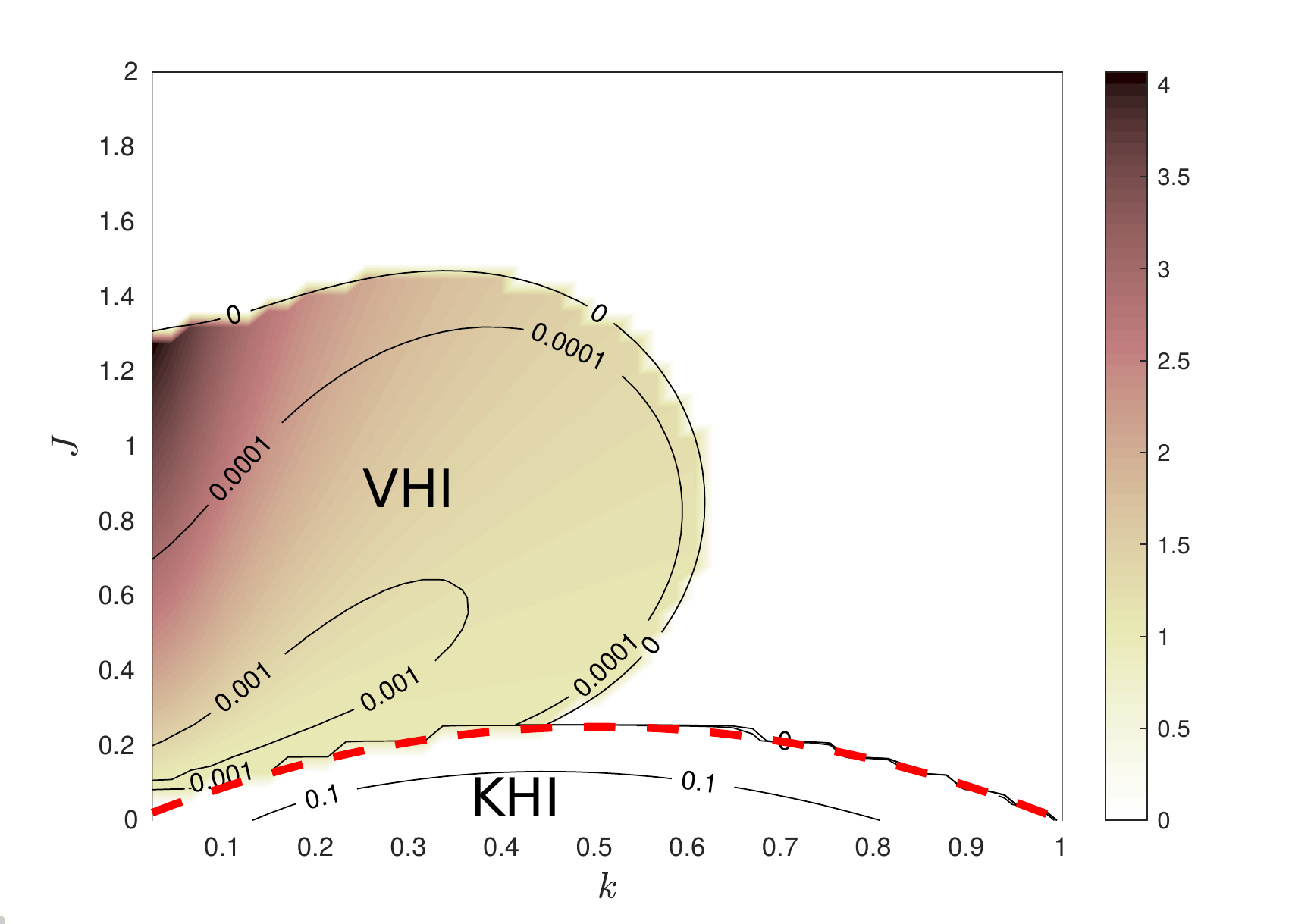}
    \caption{Stability diagram for the Hazel model flow profile
    as defined in (\ref{eq:hazel}) with
    $R=1$, $Re=500$, $Pr=1$, with boundaries at $z=\pm L_z=\pm15$.
    The contours show the growth rate of two-dimensional normal mode perturbations of wavenumber $k$, at bulk Richardson number $J$.
    The colours show the phase speed.
    The lower region, up to $J=0.25$, is KHI with zero phase speed. The upper lobe is viscous 
    Holmboe instability  (VHI),
    with non-zero phase speed.
    The dashed line shows the analytic stability boundary $J=k(1-k)$ for an unbounded domain in the inviscid limit \citep{miles_stability_1961}.
    In this, and all the 
    stability diagrams in this paper, a waviness is apparent near stability boundaries.
    This is a common problem in such stability diagrams \citep{hogg_ivey_2003, smyth_winters_2003, carpenter_2010, carpenter_2013},
    and is associated with interpolating near sharp changes of gradient in contour plots.
}
    \label{fig:basicviscousholmboe}
\end{figure}

\begin{figure}
    \centering
    \includegraphics[width=0.49\textwidth]{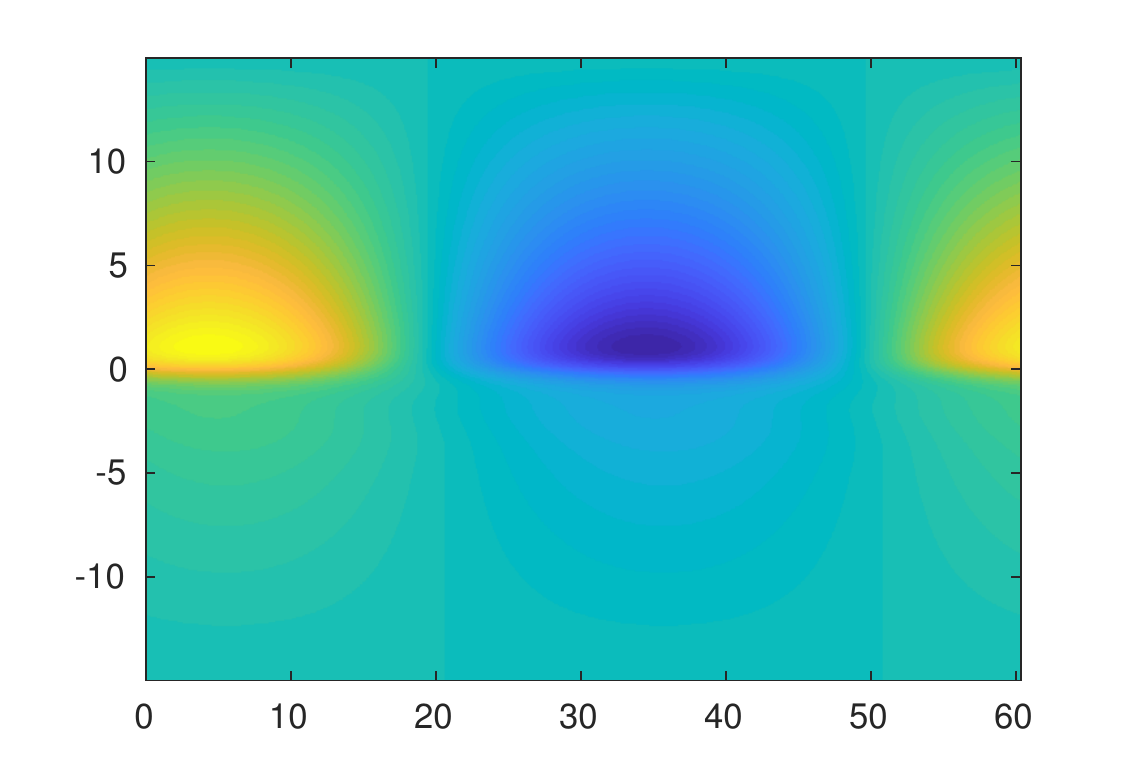}
    \includegraphics[width=0.49\textwidth]{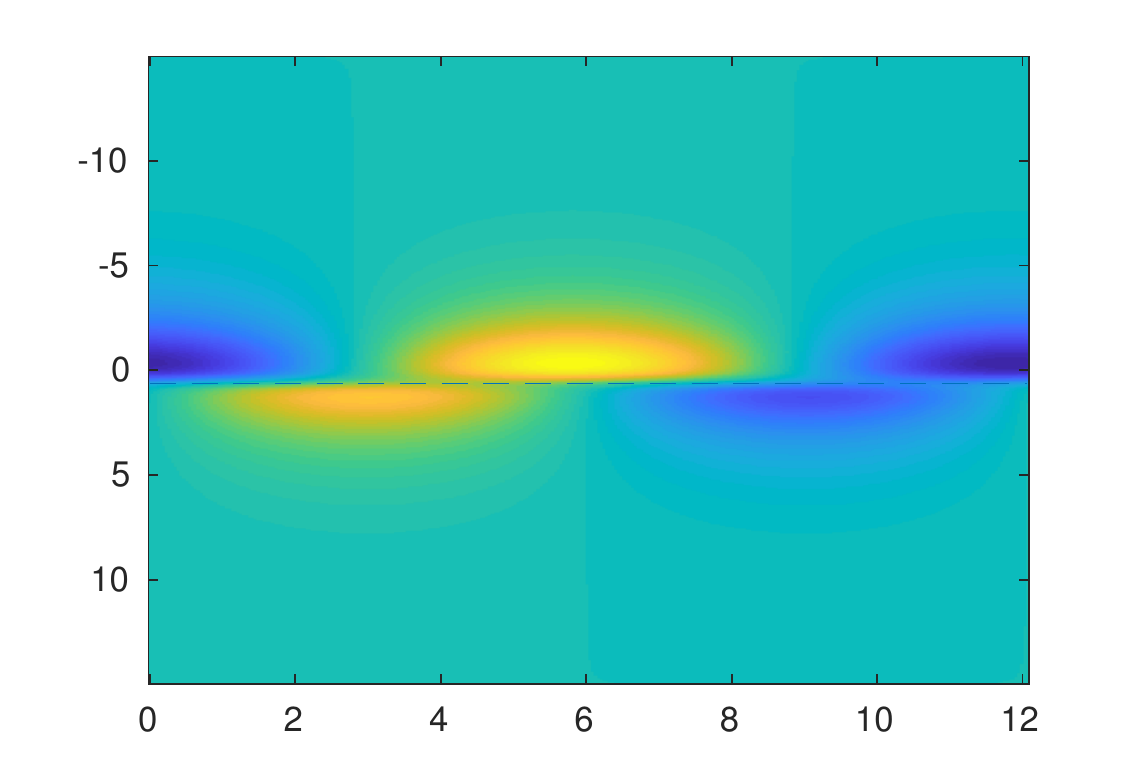}
    \caption{Vorticity field for the most unstable viscous Holmboe instability (VHI) mode for $R=1$ ($J=0.2128$, $k=0.1042$, left) and $R=3$ ($J=0.8085$, $k=0.5208$, right).
    In the latter case, a critical layer exists at $z=0.63437$ and is marked with a dashed line.}
    \label{fig:eigenfunctions}
\end{figure}

\begin{table}
    \begin{center}
\begin{tabular}{c c c c c c c c c}
    $Re$ & $R$ & $Pr$ & $L_z$ & $\sigma^{max}$ & $c_r^{max}$ & $J^{max}$ & $k^{max}$ & Figure\\[3pt]
500 & 1 & 1 & 15 & 0.002031 & 1.114 & 0.21277 & 0.10417& 1\\
500 & 1 & 1 & 10 & 0.0019489 & 1.1152 & 0.21277 & 0.0625& 3a\\
500 & 1 & 1 & 20 & 0.0020869 & 1.1363 & 0.12766 & 0.041667& 3b\\
500 & 1 & 0.7 & 15 & 0.0027958 & 1.1185 & 0.21277 & 0.10417& 4a\\
500 & 1 & 7 & 15 & 0.00056834 & 1.2536 & 0.17021 & 0.020833& 4b\\
500 & 0.5 & 0.25 & 15 & 0.0003781 & 1.557 & 0.12766 & 0.041667& 5a\\
500 & 1.5 & 2.25 & 15 & 0.0032918 & 1.0156 & 0.29787 & 0.10417& 5b\\
500 & 2 & 4 & 15 & 0.0033963 & 0.88215 & 0.38298 & 0.125& 5c\\
500 & 3 & 9 & 15 & 0.031314 & 0.56129 & 0.80851 & 0.52083& 5d\\
5.5 & 1 & 1 & 15 & 0.0014049 & 1.4289 & 0.29787 & 0.125& 6\\
6 & 1 & 1 & 15 & 0.0025824 & 1.2608 & 0.21277 & 0.125& 6\\
7 & 1 & 1 & 15 & 0.0039878 & 1.2708 & 0.25532 & 0.14583& 6\\
10 & 1 & 1 & 15 & 0.0067811 & 1.2525 & 0.25532 & 0.14583& 6\\
15 & 1 & 1 & 15 & 0.0092699 & 1.2498 & 0.34043 & 0.1875& 6\\
20 & 1 & 1 & 15 & 0.01023 & 1.2336 & 0.34043 & 0.1875& 6\\
25 & 1 & 1 & 15 & 0.010546 & 1.2515 & 0.34194 & 0.175& 6\\
30 & 1 & 1 & 15 & 0.010542 & 1.2334 & 0.35806 & 0.1875& 6\\
40 & 1 & 1 & 15 & 0.01 & 1.2446 & 0.34043 & 0.16667& 6\\
100 & 1 & 1 & 15 & 0.0069558 & 1.1762 & 0.25532 & 0.125& 6\\
200 & 1 & 1 & 15 & 0.0044792 & 1.1263 & 0.21277 & 0.10417& 6\\
400 & 1 & 1 & 15 & 0.0025701 & 1.1279 & 0.17021 & 0.0625& 6\\
1000 & 1 & 1 & 15 & 0.0011049 & 1.1183 & 0.17021 & 0.0625& 6\\
2000 & 1 & 1 & 15 & 0.00057664 & 1.0722 & 0.13226 & 0.025& 6\\
4000 & 1 & 1 & 15 & 0.0002994 & 1.0854 & 0.13226 & 0.0125& 6\\
10000 & 1 & 1 & 15 & 0.00012168 & 1.0827 & 0.13226 & 0.0125& 6\\
\end{tabular}
\caption{The various parameters used for the linear stability diagrams, as well as the maximum growth rate $\sigma^{max}$
of viscous Holmboe instability (VHI)
         for each set of parameters, and the phase speed $c_r^{max}$, wavenumber $k^{max}$ and bulk Richardson number $J^{max}$
         at which they occur.}
\label{tab:results}
\end{center}
\end{table}

Figure \ref{fig:basicviscousholmboe} shows a typical example of the viscous Holmboe instability (VHI).
There is a clear distinction between those unstable modes with zero phase speed, which we identify as KHI,
and the modes with non-zero phase speed, which we identify as VHI.
Though the existence of unstable modes at $R=1$ with non-zero phase speed was unknown before \citet{pck19}, the diagram bears a striking
resemblance to the classic stability diagrams for inviscid HWI for a piecewise linear profile with a
density discontinuity \citep[figure 7]{holmboe_1962} and the Holmboe model with $R>2$ \citep[figure 8]{hazel1972numerical}.
Crucially, above $J=0.25$ on this diagram, the gradient Richardson number of the flow is everywhere 
greater
than one quarter, and so we expect stability as $Re\to\infty$.
In the inviscid case, as $J$ is increased the dominant KHI mode and a subdominant KHI mode converge
and bifurcate into the pair of HWI modes, with opposite phase speeds.
In the viscous case, the regions of the two instabilities overlap slightly and there is no clean bifurcation
from one to the other.
The remainder of this section will
explore how 
the structure of stability  diagrams like figure \ref{fig:basicviscousholmboe}
change as various parameters are varied.

Figure \ref{fig:eigenfunctions} shows typical eigenmodes of the spanwise vorticity.
With $R=1$, i.e. a density interface as wide as the shear layer, no critical layer exists.
With $R=3$, the eigenmode is virtually indistinguishable from the $Re\to\infty$ case, and a critical layer
is present and clearly manifests itself within the spatial structure of the mode.
Both of these modes have an equivalent mode associated with the complex conjugate eigenvalue,
which is identical except for a reflection in the centreline.
In the $R=1$ case, we also note that the growth rate is maximised at a much lower wavenumber.

Table \ref{tab:results} shows the full list of parameters for which stability diagrams were produced.
For each diagram, we find the maximum growth rate for  
VHI, i.e. the maximum of $\sigma$ such that the phase speed $c_r$ is non-zero, maximised over the discretised
values of $k$ and $J$. Since the grids are relatively coarse, the values will not be the true maxima
as no optimisation algorithm has been employed, but they give a strong indication of the trend.


\subsection{Effects of domain height}
\label{sec:l}

\begin{figure}
    \centering
    \includegraphics[width=0.49\textwidth]{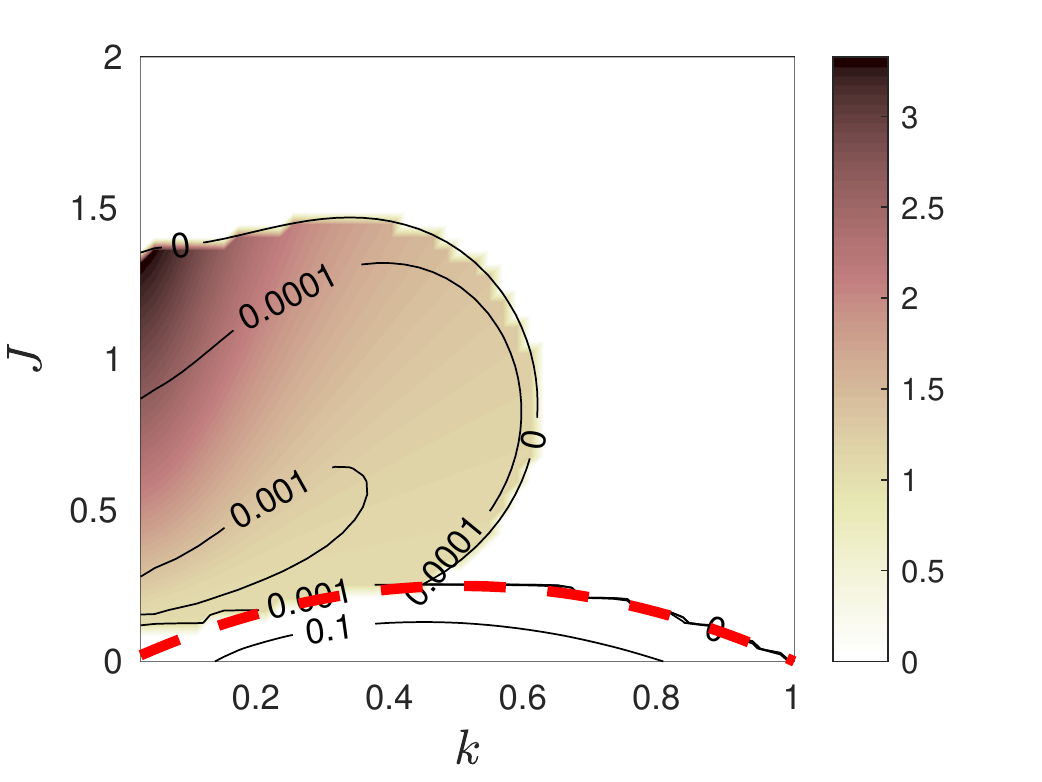}
    \includegraphics[width=0.49\textwidth]{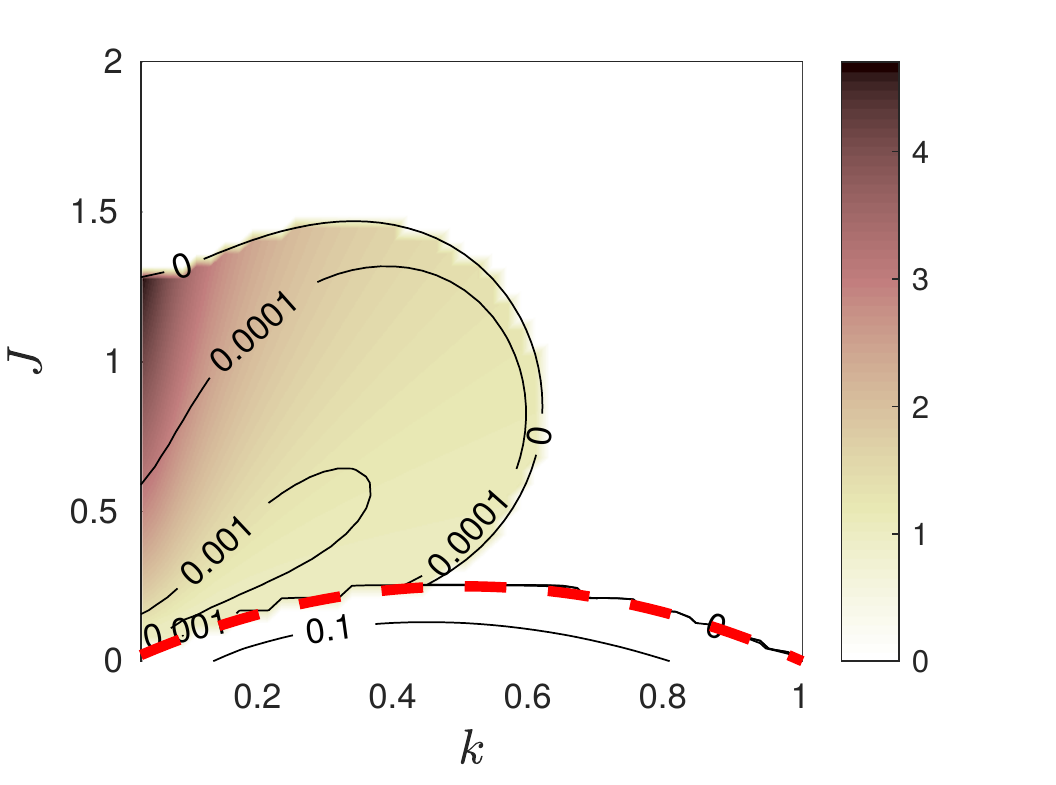}
    \caption{As for figure \ref{fig:basicviscousholmboe}, but with $L_z=10$ (left) and $L_z=20$ (right).}
    \label{fig:varyheight}
\end{figure}

The instabilities we study, KHI and HWI, were originally derived as solutions to the Taylor-Goldstein
equation in an unbounded domain. There are several ways to approximate a domain of infinite height numerically,
but we choose the simplest, which is to use a domain of sufficiently large, but finite, height.
How large is sufficient is an important question, as a very large domain is computationally inefficient.
Certainly as the height gets small compared with the wavelength of the instabilities we expect
the results to change dramatically, and
\citet{hazel1972numerical} noted how the diagrams always differ from the analytic, unbounded results at
low wavenumbers.
Figure \ref{fig:varyheight} shows the same diagram as figure \ref{fig:basicviscousholmboe},
but at a larger and smaller domain heights. Though the results are slightly different, qualitatively they
are very similar, especially for $L_z=20$, with $L_z=10$ showing more instability at low wavenumbers.
The maximum growth rate of the VHI region is $\sigma=1.9489\times10^{-3}$ for $L_z=10$ and $\sigma=2.0869\times10^{-3}$ for $L_z=20$,
compared with $\sigma=2.0310\times10^{-3}$ for $L_z=15$,
suggesting that $L_z=15$ is sufficient to capture
the behaviour in which we are interested.

\subsection{Effects of Prandtl number}
\label{sec:pr}

\begin{figure}
    \includegraphics[width=0.49\textwidth]{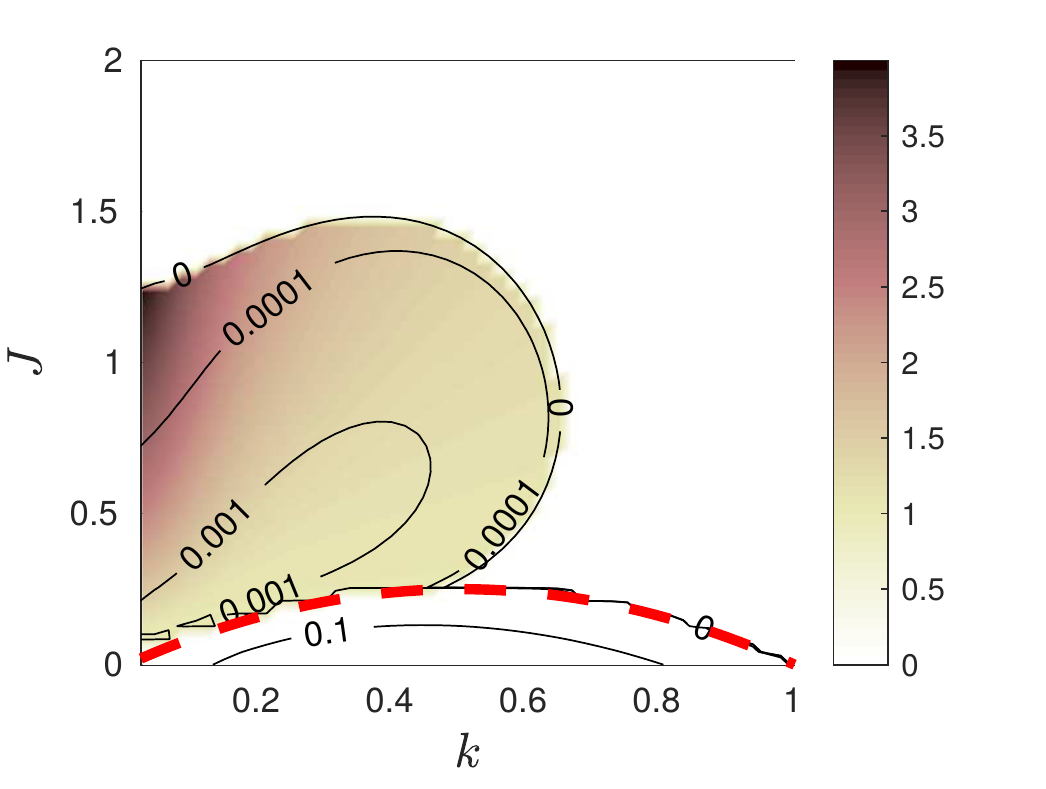}
    \includegraphics[width=0.49\textwidth]{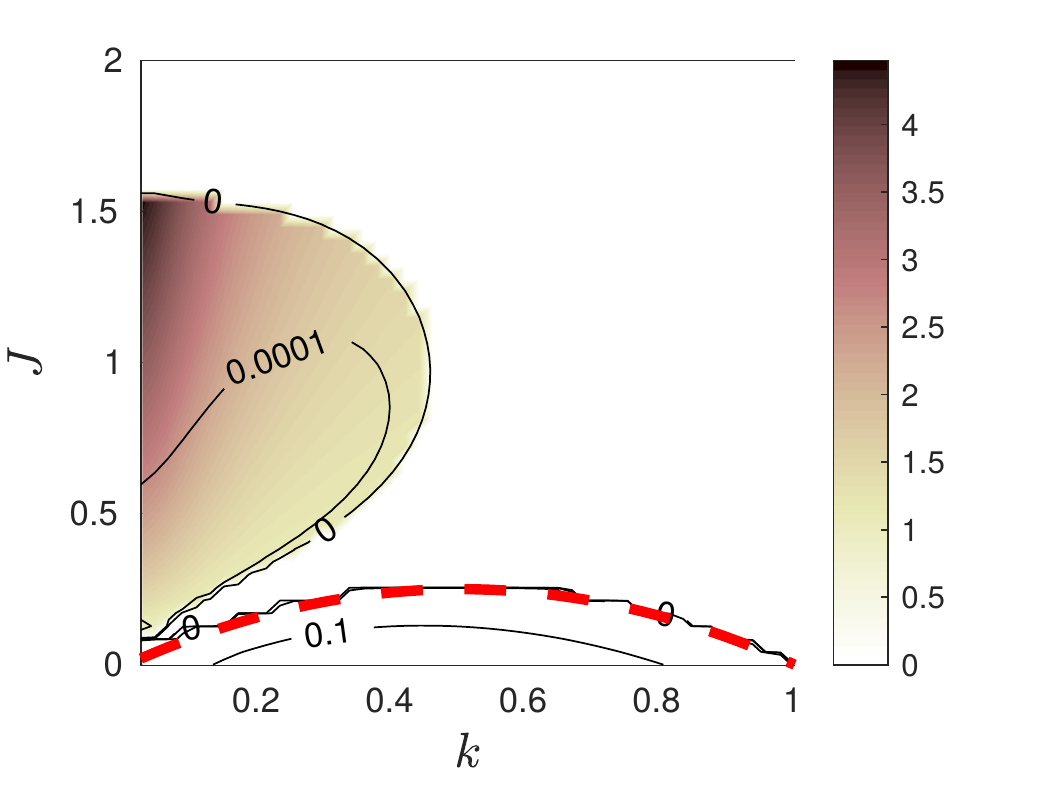}
    \caption{As for figure \ref{fig:basicviscousholmboe}, but with $Pr=0.7$ (left) and $Pr=7$ (right).}
    \label{fig:varypr}
\end{figure}

Figure \ref{fig:varypr} shows the effect on the stability diagram of varying the Prandtl number.
For $Pr=0.7$ (characteristic of thermally-stratified air), we find a maximum growth rate of $\sigma=2.7958\times10^{-3}$, and for $Pr=7$ (a typical value for thermally-stratified water) of $\sigma=5.6834\times10^{-4}$,
compared with $\sigma=2.0310\times10^{-3}$ for $Pr=1$. Therefore, decreasing the diffusion of buoyancy
seems to have a stabilising effect on  
VHI.
In contrast,
the KHI at the bottom of the figure is virtually unchanged as $Pr$ is varied by an order of magnitude, which reinforces the
idea that KHI is produced by the shear alone.
\citet{jones1977} found strong instability at very low $Pr$, but we believe this to be a different effect.

Henceforth we give results using $Pr=R^2$, as proposed by \citet{smyth_klaasen_peltier}.
Despite the fact that the instability seems to be destabilised when $Pr$ is reduced, it is also
stabilised when $R$ is decreased, as we shall see.

\subsection{Effects of $R$}
\label{sec:r}

\begin{figure}
    \begin{subfigure}{0.5\linewidth}
        \includegraphics[width=\textwidth]{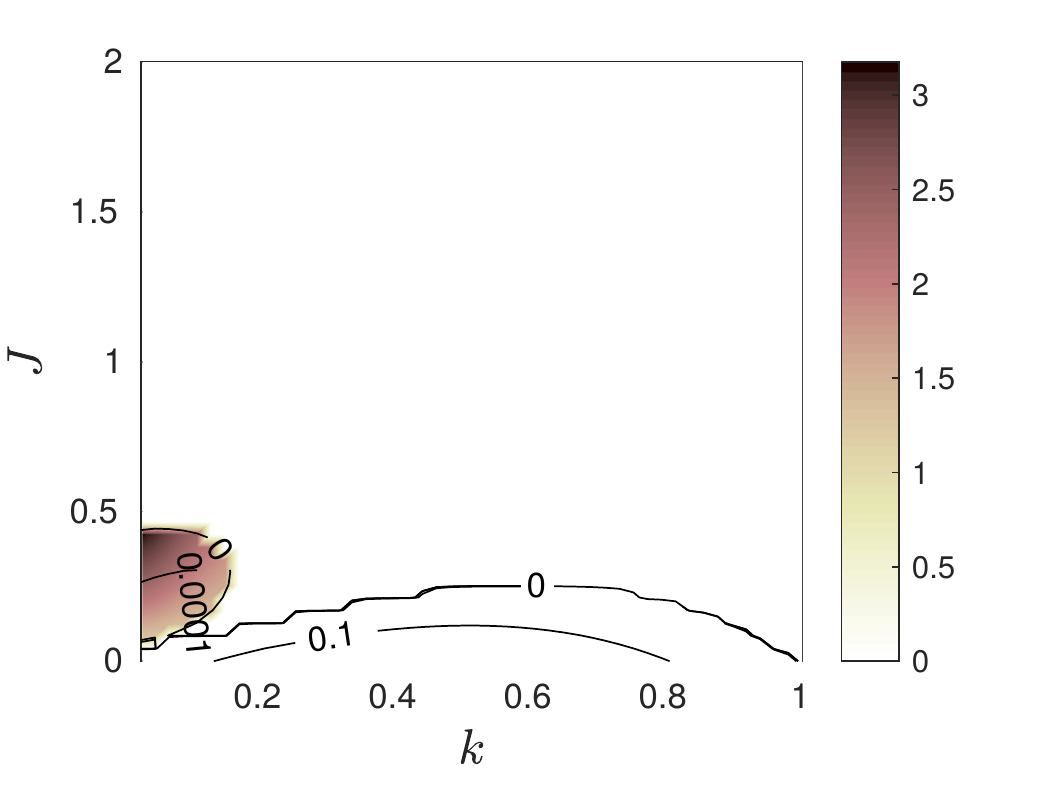}
        \caption{}
    \end{subfigure}
    \begin{subfigure}{0.5\linewidth}
        \includegraphics[width=\textwidth]{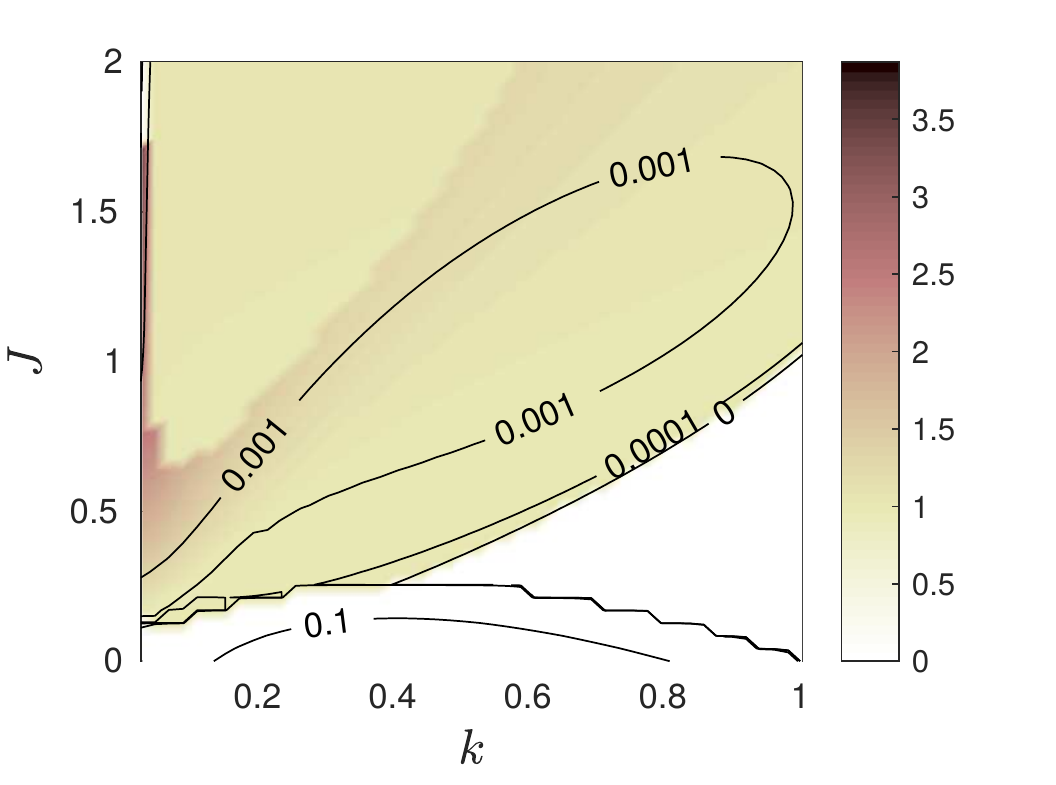}
        \caption{}
    \end{subfigure}
    \begin{subfigure}{0.5\linewidth}
        \includegraphics[width=\textwidth]{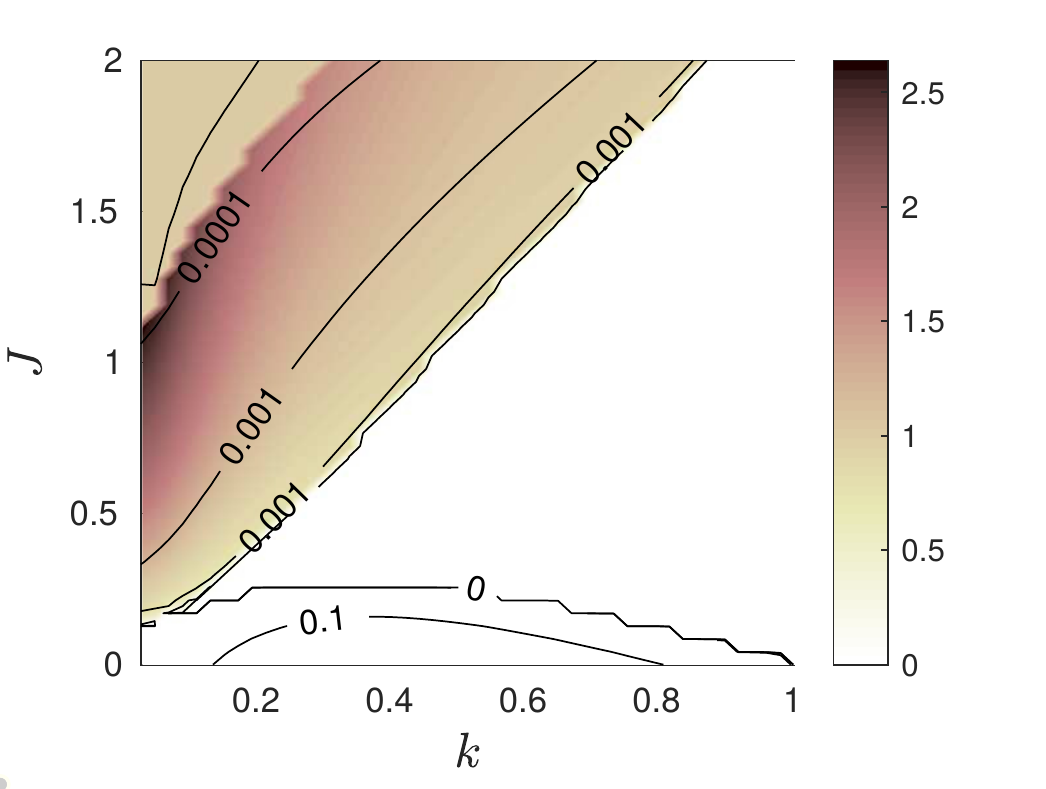}
        \caption{}
    \end{subfigure}
    \begin{subfigure}{0.5\linewidth}
        \includegraphics[width=\textwidth]{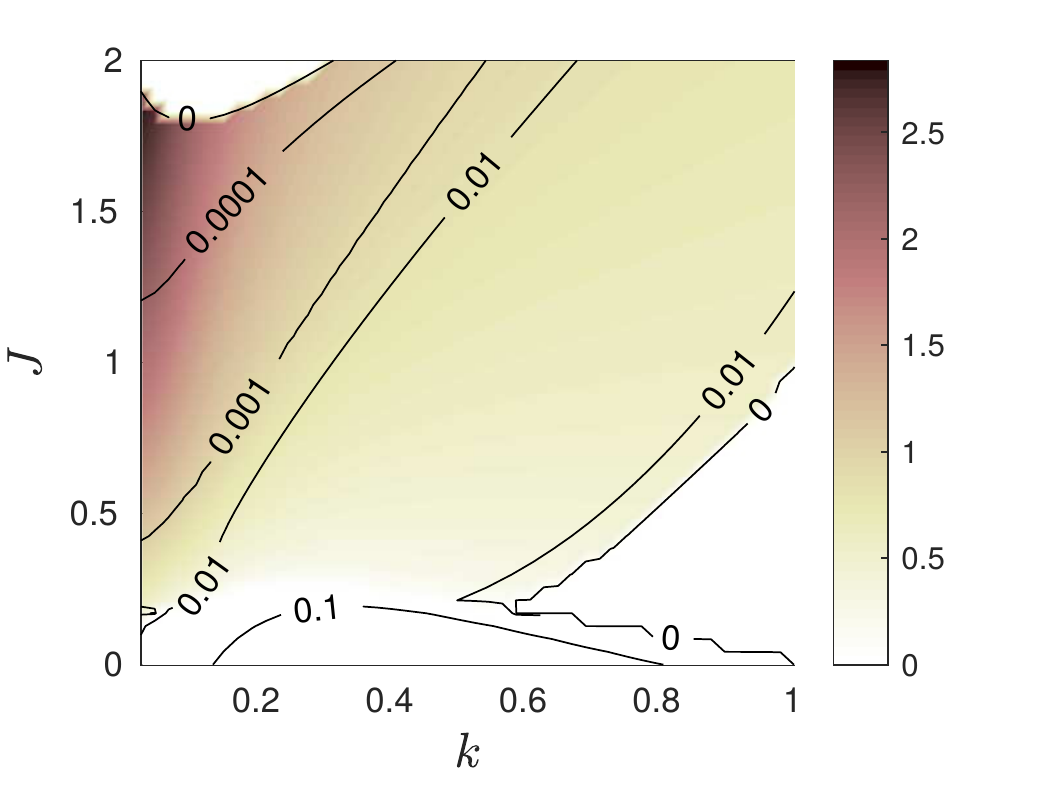}
        \caption{}
    \end{subfigure}

    \caption{As for figure \ref{fig:basicviscousholmboe}, but with (a) $R=0.5$, $Pr=0.25$, (b) $R=1.5$, $Pr=2.25$, (c) $R=2$, $Pr=4$, (d) $R=3$, $Pr=9$.
    Only the last of these would exhibit HWI at $Re=\infty$.}
    \label{fig:varyr}
\end{figure}

So far, the results we have presented have concentrated on $R=1$, the original Holmboe model.
However, in the inviscid limit, HWI exists only for $R>2$ \citep{alexakis2007}.
Figure \ref{fig:varyr} shows the stability diagram at $Re=500$ over a range of $R$, with $Pr=R^2$.
All diagrams show a region of instability with non-zero phase speed, which we identify as VHI.
In the case $R=3$, the diagram is very similar to the classical diagram of an inviscid fluid \citep{hazel1972numerical}.
The unstable region above the usual band, at low wavenumbers, has $c_r>1$, so there  is no critical layer.
As $R\to2$ from above, the inviscid results suggest that the band should narrow to a line \citep{alexakis2005holmboe},
but instead we see a significant region of instability. In the diagrams for $R=1.5$ and $R=2$, a second band of instability
is observed above the first, with reduced phase speed, and we conjecture that this may be connected
with the higher Holmboe modes. This has not been investigated further, as the growth rate here is vanishingly small.

In all cases, though it is not clear from the truncated diagrams, the instability is suppressed at large $k$ by viscosity.
This is in contrast to the inviscid limit, which has instability at arbitrarily large $k$ and $J$.
It is only in this large $k$ limit that the wave interaction arguments can be made rigorous.

\subsection{Effects of Reynolds number}
\label{sec:re}

The Miles-Howard theorem tells us that VHI at $R=1$ 
must disappear for
$J>1/4$,
in the inviscid limit $Re\to\infty$. This leaves many possibilities: 1) the region of instability could retreat below
$J=1/4$; 2) the region could shrink; or 3) the growth rates could vanish but the region remain a constant size.
There may or may not be some finite $Re$
above which VHI does not exist. It is also important to ask at what value of $Re$ the growth of the instability is the fastest,
or indeed the relative growth rate compared with the diffusion of the background profile.

The growth rate is maximised between $Re=25$ and $Re=30$, with value $\sigma\approx 0.0105$.
The relative growth rate $\sigma Re$, which is required to be large compared with unity for a physically relevant instability,
was found to increase with $Re$ (at least until $Re=10000$), which is a curious result, since it means that despite the growth rate being
maximised at a very low value of $Re$, in practice we are more likely to observe the instability at much higher $Re$.

\begin{figure}
    \centering
    \includegraphics[width=0.9\textwidth]{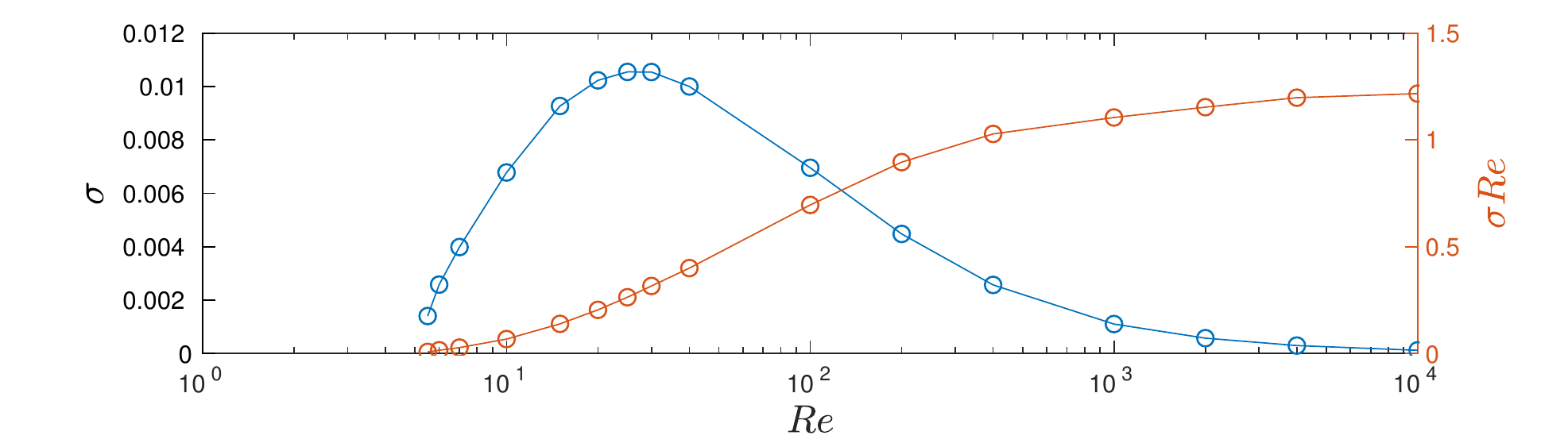}
    \caption{The growth rate $\sigma$ (left axis) and relative growth rate $\sigma Re$ (right axis), maximised over $k$ and $J$, for VHI at $R=1$, $Pr=1$, as $Re$ varies.}
    \label{fig:growthagainstre}
\end{figure}

The critical Reynolds number $Re_c$ for the viscous Holmboe instability at $R=1$, $Pr=1$, below
which there is no instability except KHI, was found to be $Re=4.615$.
At criticality, the instability appears at $J=0.25$ and $k=0.12$.
This is in contrast with KHI, which at $J=0$ was found to have $Re_c = 0$
\citep{betchov1963stability}. In that case, viscosity has a purely stabilising effect.

\section{Nonlinear evolution}
\label{sec:nonlinear}
\citet{smyth_peltier_1990} showed that at low Reynolds numbers, the linear evolution of HWI is
insufficiently fast to overcome the diffusion of the background flow. This leads to the possibility
that VHI, for which the growth rates are always small, never physically
manifests when the background flow is allowed to diffuse.
We consider the nonlinear evolution, which allows us to see whether the viscous Holmboe instability
develops the classic counter-propagating vortices of HWI.
We use the same DNS code as \citet{pck19} to solve the full Boussinesq equations, which is pseudospectral in the streamwise direction
and utilises finite differences in the vertical. In the present case, the background flow is allowed to diffuse.

\begin{figure}
    \begin{subfigure}{0.5\linewidth}
        \includegraphics[width=\textwidth]{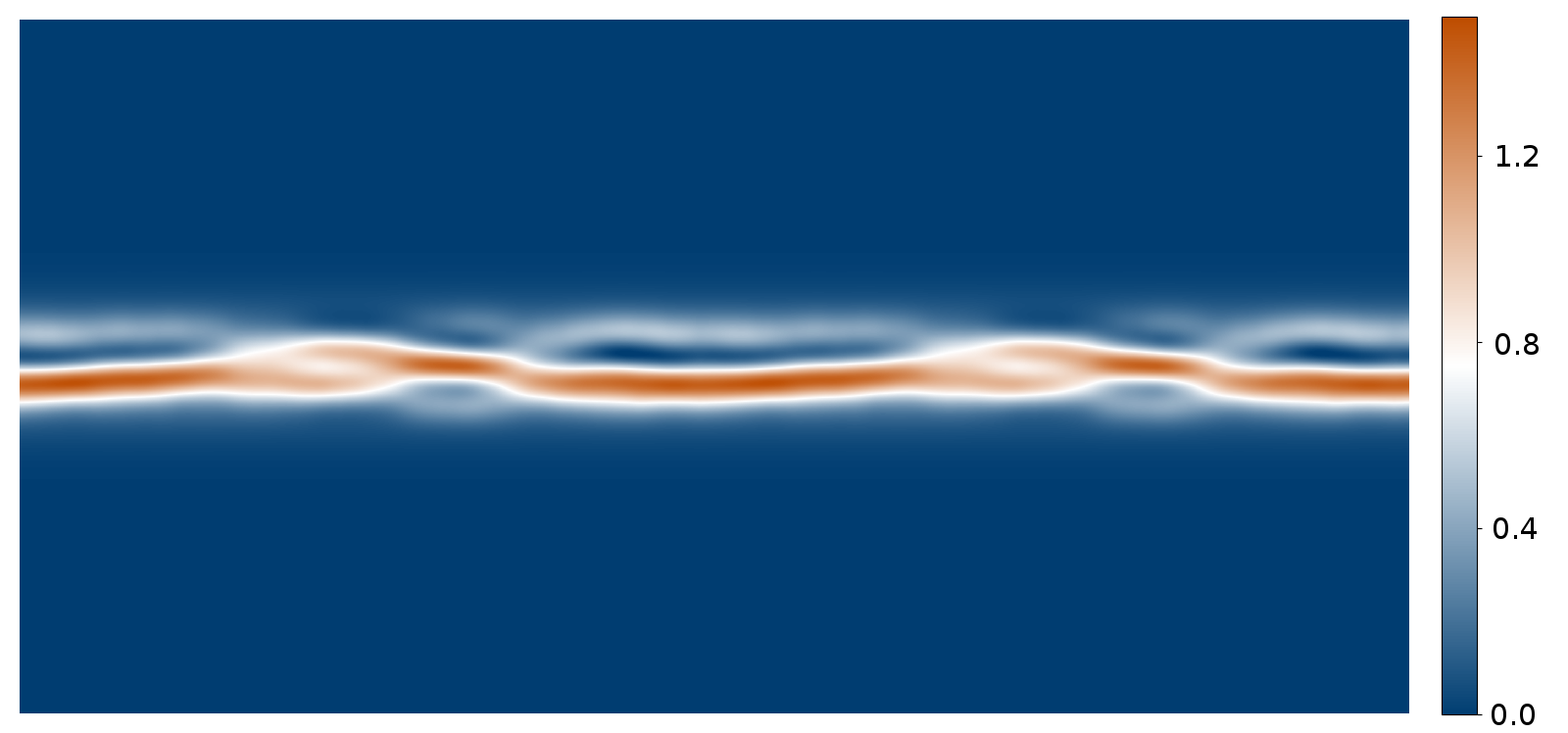}
        \caption{}
    \end{subfigure}
    \begin{subfigure}{0.5\linewidth}
        \includegraphics[width=\textwidth]{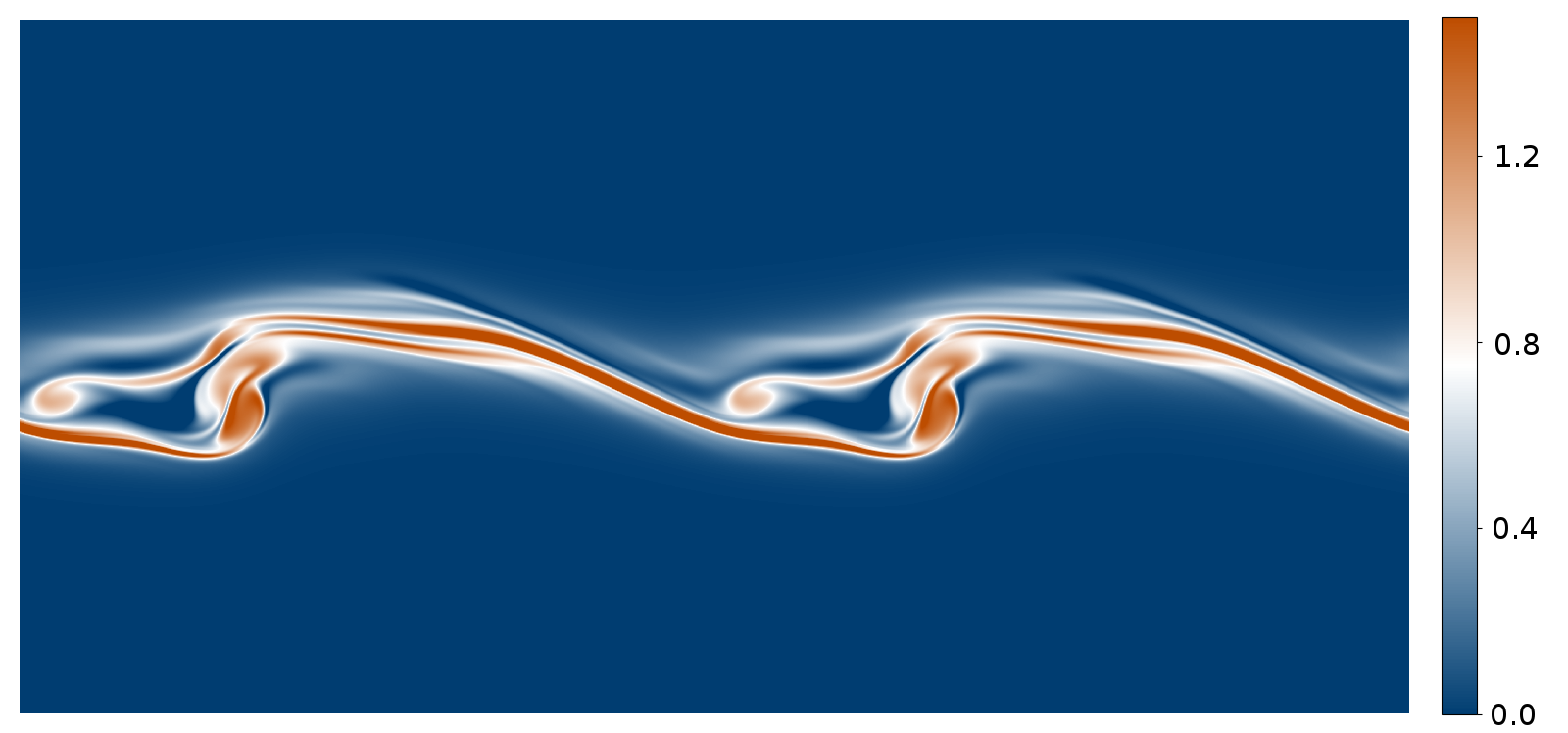}
        \caption{}
    \end{subfigure}
    \begin{subfigure}{0.5\linewidth}
        \includegraphics[width=\textwidth]{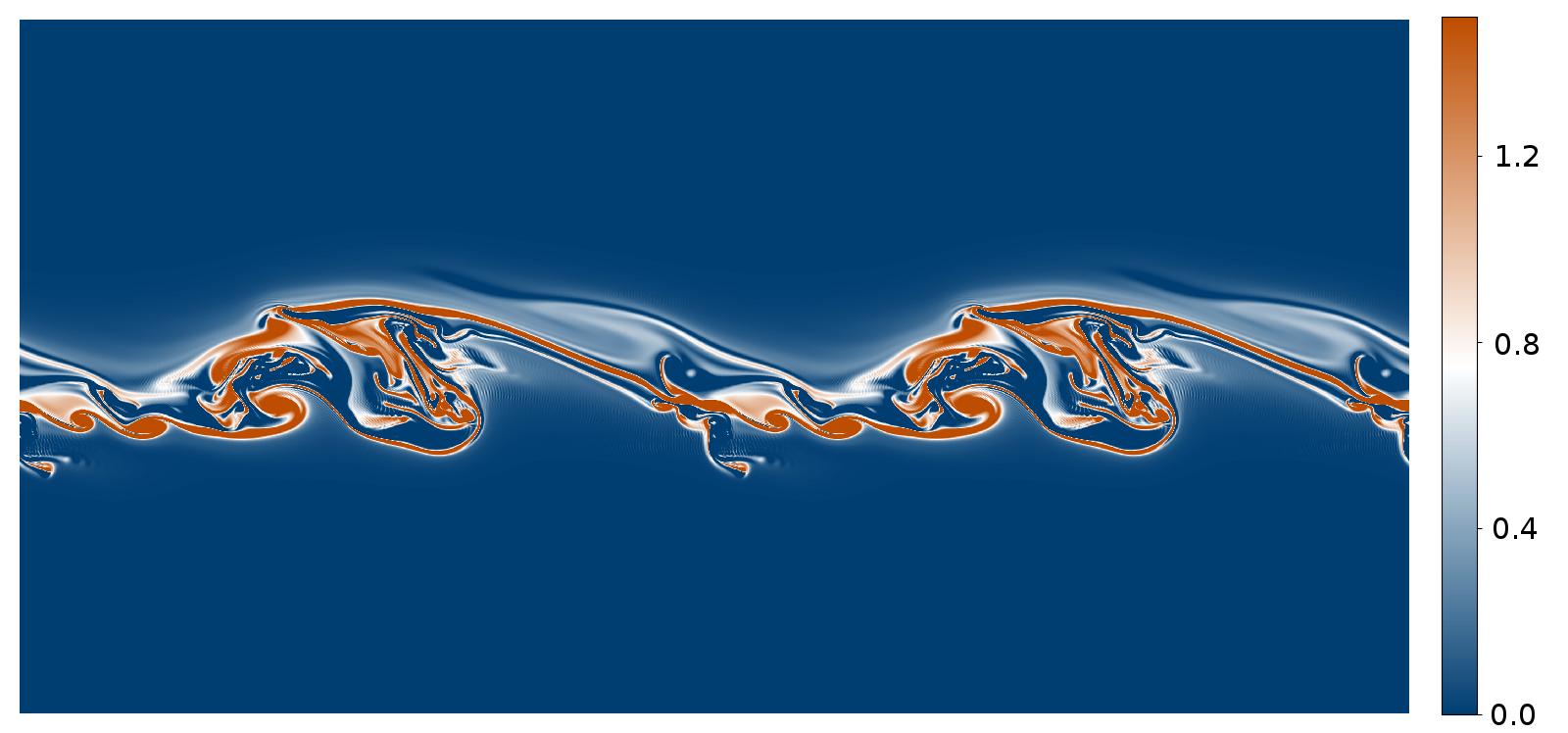}
        \caption{}
    \end{subfigure}
    \begin{subfigure}{0.5\linewidth}
        \includegraphics[width=\textwidth]{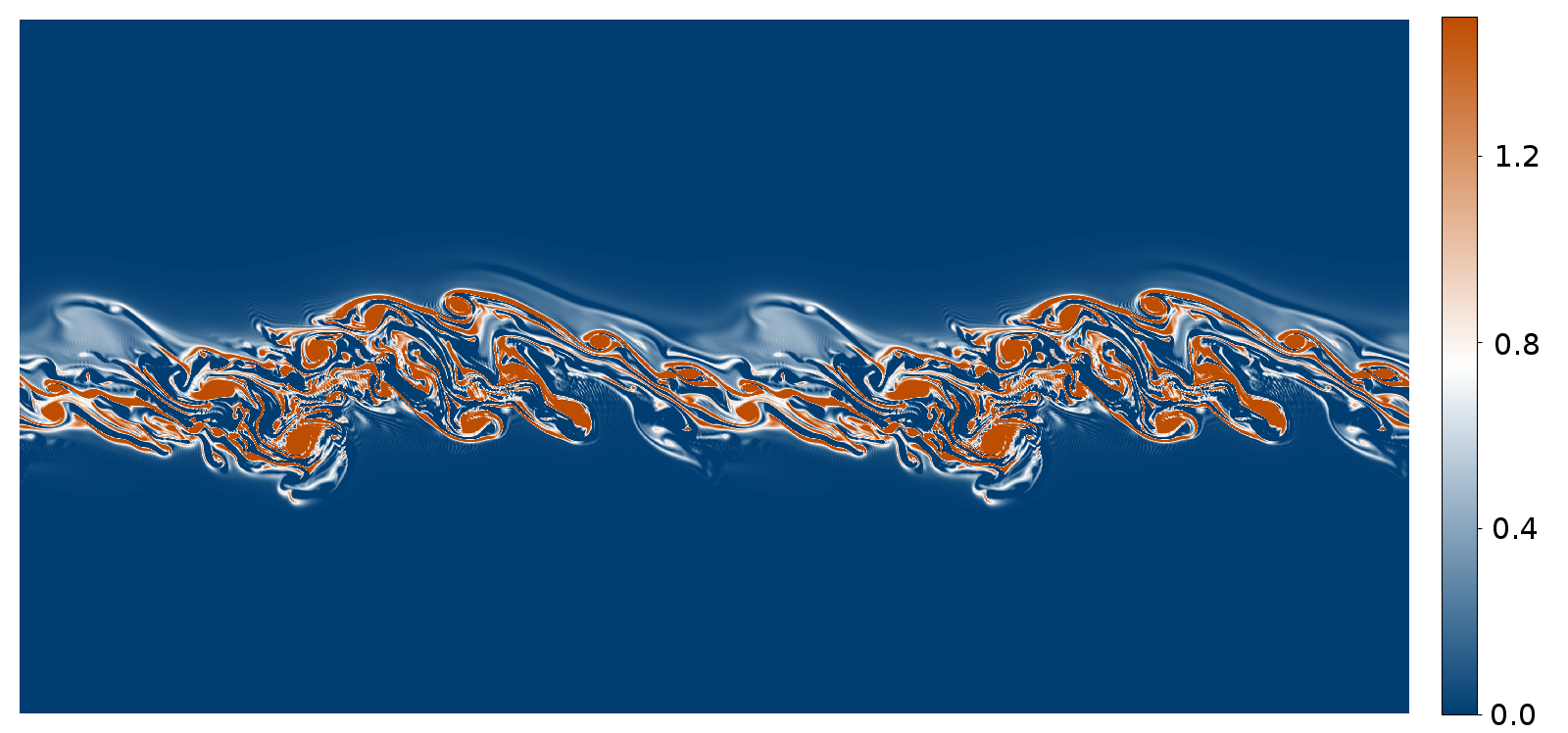}
        \caption{}
    \end{subfigure}

    \caption{The total vorticity field of a two-dimensional nonlinear simulation of the Boussinesq equations
             at $Re=4000$, $Pr=2.25$, $L_x=20$, $L_z=10$ and $J=0.1$.
             The initial state is a background field with $R=1.5$,
             plus a perturbation of random noise in the first sixth of the horizontal Fourier modes, and the first five
             Hermite polynomials in the vertical.
             Two domain widths are shown horizontally.
             (a) $t=0$, showing the random initial conditions. (b) $t=20$, showing the Kelvin-Helmholtz billow that has begun to develop.
             (c) $t=40$, showing that the billow has saturated and is starting to break down. (d) $t=60$, showing that the KHI has led to (two-dimensional) turbulence. An animation  of the evolving flow is available as  supplementary material.}
    \label{fig:nonlinearkh}
\end{figure}

Here we present the results of two direct numerical simulations (restricted to two dimensions) with $R=1.5$, a case for which
no HWI is predicted in the inviscid limit. We take $Re=4000$, a compromise between maximising
the relative growth rate (see section \ref{sec:re}) and minimising the spatial resolution.
We chose a domain width of $L_x=20$, which permits multiple unstable modes.
Figure \ref{fig:nonlinearkh} shows the results of a calculation with $J=0.1$, for which we expect a Kelvin-Helmholtz instability to develop to finite amplitude.
A linear stability analysis 
predicts exponential growth rates of $\sigma=0.1244$ and $\sigma=0.0981$ for mode 1 and mode 2 disturbances ($k=\pi/10$ and $k=\pi/5$) respectively,
in both cases with zero phase speed.
We use a relatively large initial perturbation of random noise in low wavenumber Fourier and Hermite modes,
which, along with the comparable growth rates for the two unstable modes, leads to an incoherent, but nevertheless recognisable, Kelvin-Helmholtz billow.
At the large $Re$ studied, this rapidly breaks down into turbulence, and significant mixing is achieved, although it is important to remember that this DNS is restricted to two dimensions, and so the specific characteristics of the mixing are likely to be unphysical.

\begin{figure}
    \begin{subfigure}{0.5\linewidth}
        \includegraphics[width=\textwidth]{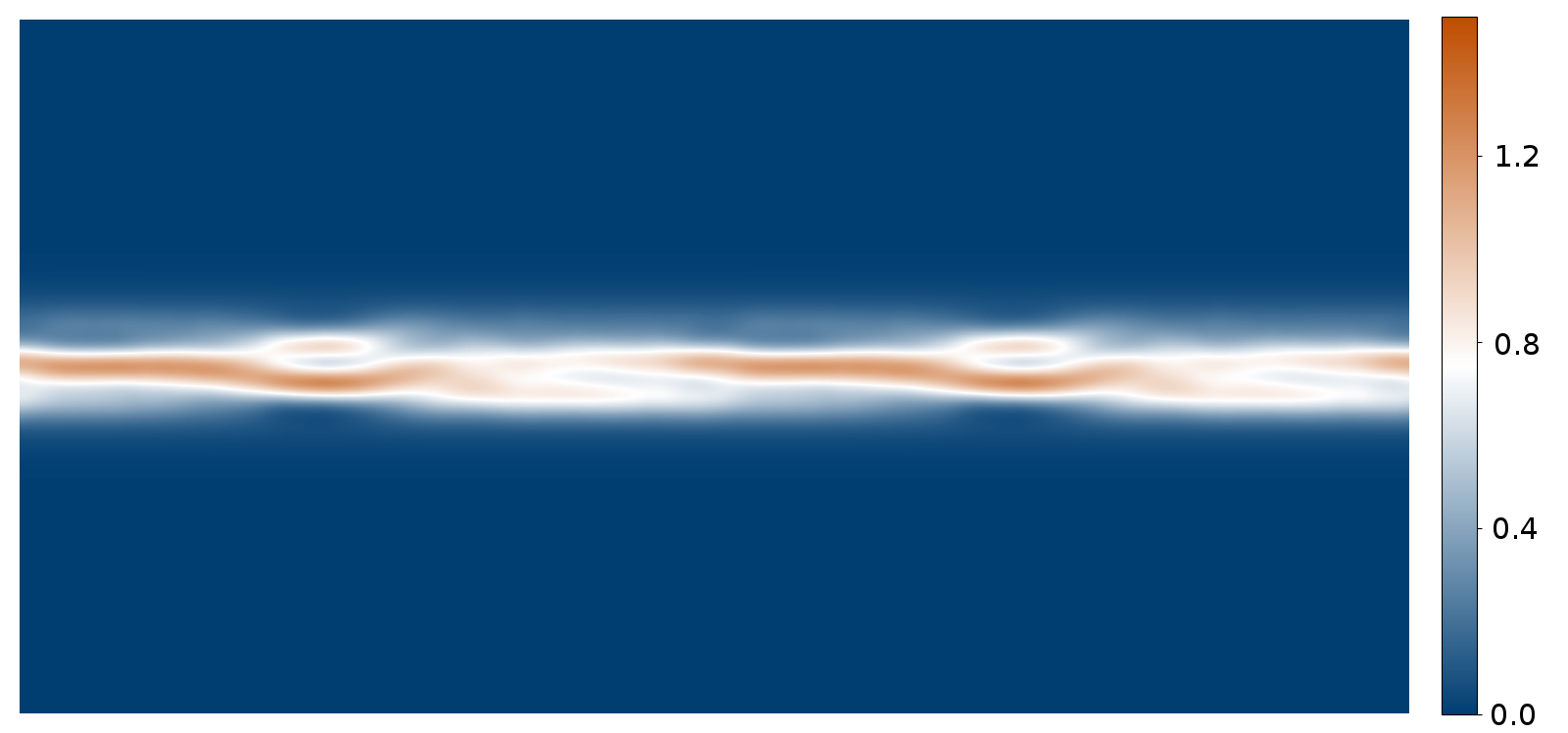}
        \caption{}
    \end{subfigure}
    \begin{subfigure}{0.5\linewidth}
        \includegraphics[width=\textwidth]{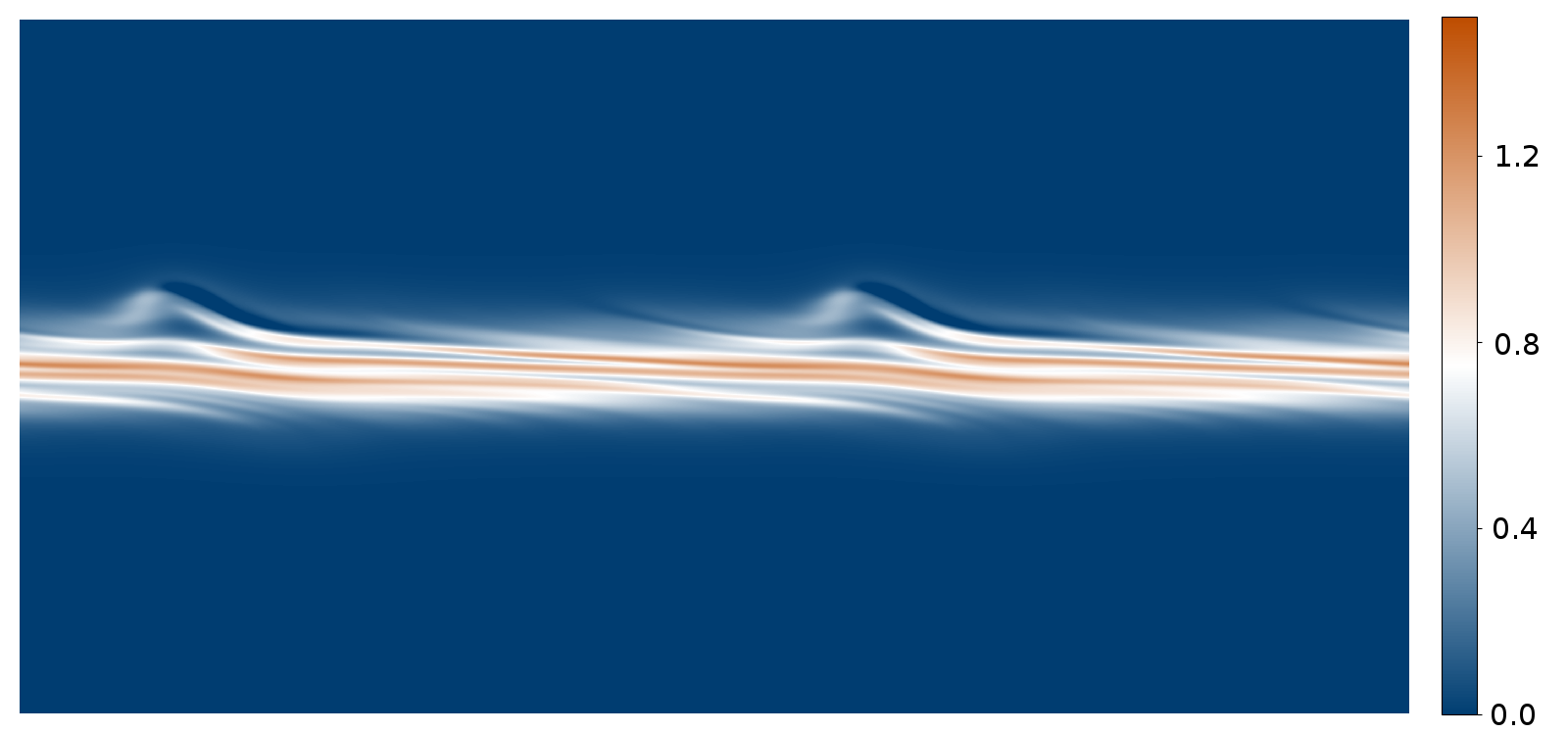}
        \caption{}
    \end{subfigure}
    \begin{subfigure}{0.5\linewidth}
        \includegraphics[width=\textwidth]{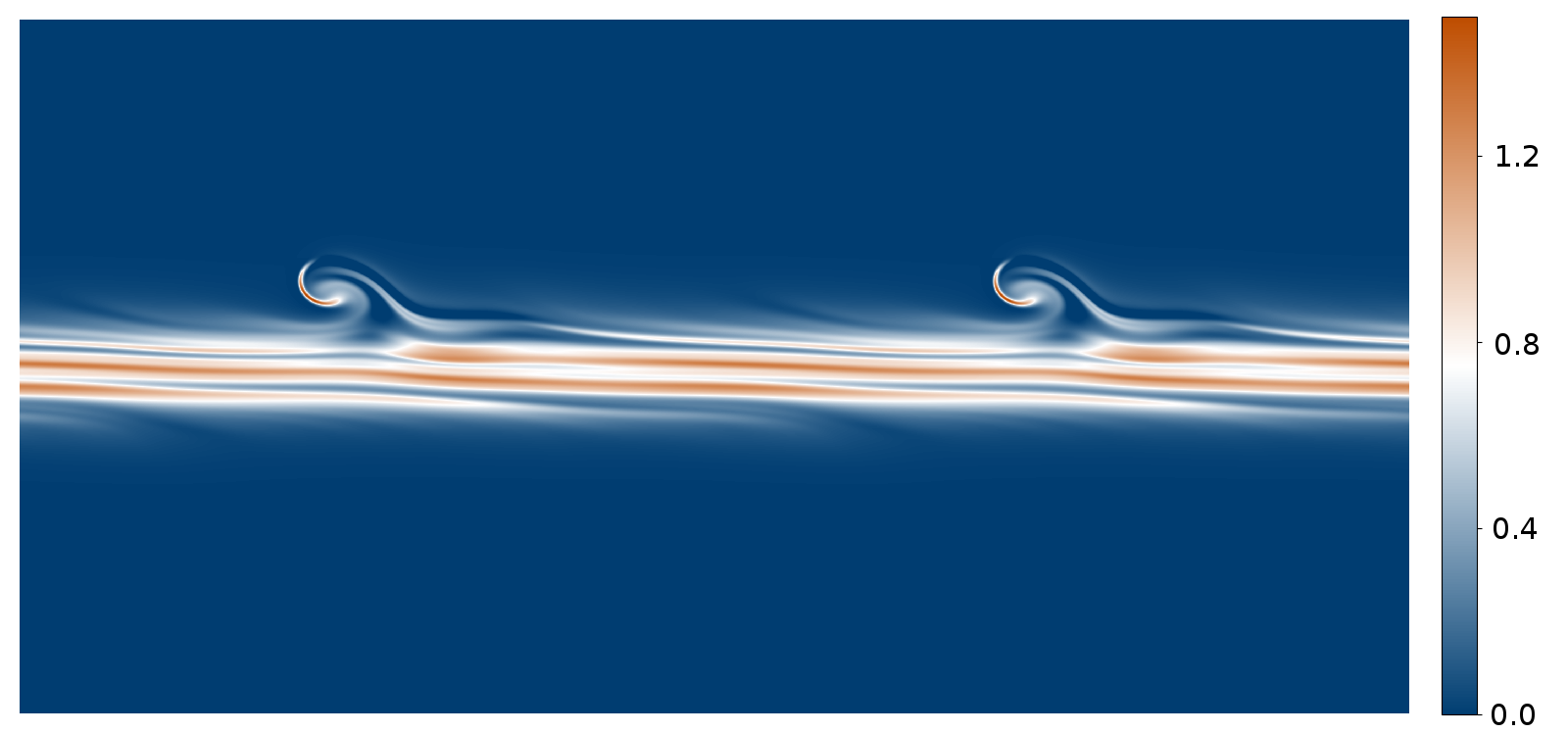}
        \caption{}
    \end{subfigure}
    \begin{subfigure}{0.5\linewidth}
        \includegraphics[width=\textwidth]{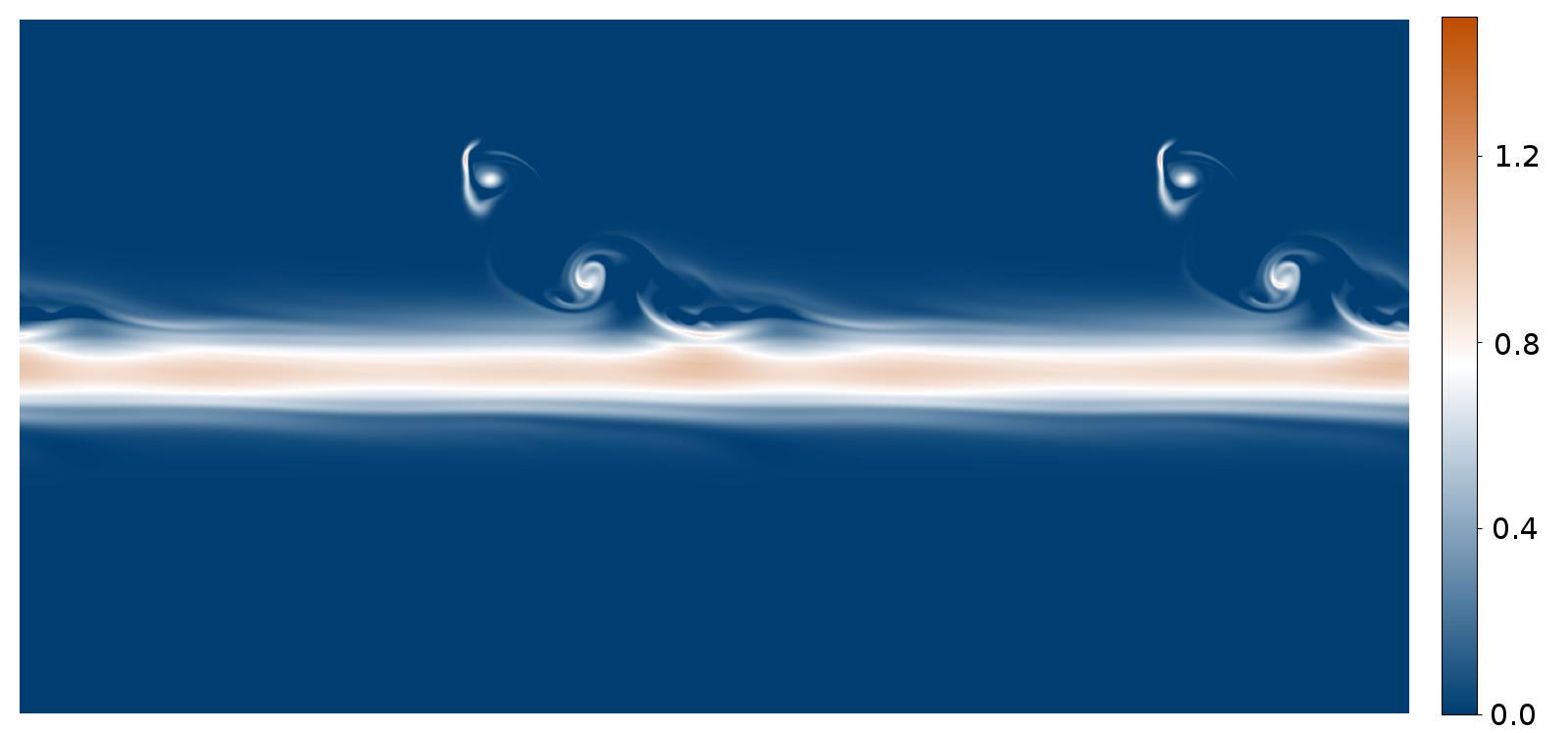}
        \caption{}
    \end{subfigure}

    \caption{As for figure \ref{fig:nonlinearkh}, but with $J=0.4355$.
             (a) $t=0$, showing the random initial conditions.
             (b) $t=20$, showing  that a `cusped wave' is apparent, characteristic of HWI at finite amplitude. (c) $t=35$, showing that a leftwards-propagating vortex is now visible above the shear layer.
             (d) $t=110$, showing that the vortex has weakened as the mixing layer diffuses away. An animation  of the evolving flow is available as  supplementary material.}
    \label{fig:nonlinearvh}
\end{figure}

Figure \ref{fig:nonlinearvh} shows the same calculation with $J=0.67$, which maximises the growth rate
for VHI at this wavenumber. Again, both modes 1 and 2 are unstable, with growth rates $\sigma=4.1121\times 10^{-4}$ and $\sigma=1.7012\times 10^{-4}$ respectively,
and phase velocities $c_r=\pm 1.0211$ and $c_r=\pm 1.0056$.
Since the phase speeds are greater than $1$, no critical layer exists for these instabilities.
In this case, the relative growth rate clearly does not satisfy $\sigma Re\gg 1$, so we require
a large initial perturbation to trigger significant instability. The strong asymmetry of this random
perturbation means that a Holmboe `wave' is apparent only on one side of the interface.
Despite the lack of a critical layer, a `cusped wave' very reminiscent of classic HWI \citep{alexakis2009stratified,salehipour_caulfield_peltier_2016}
is apparent, and grows large enough for a clear vortex to be apparent.
This vortex is responsible for some mixing, which can be observed when comparing the long time vorticity distribution
above the interface, where the vortex exists, to below, where no strong VHI was triggered.
However, this mixing is relatively weak compared with the diffusion of the background profile.
It is difficult to define a speed precisely for the nonlinear wave, but it appears to be close to $1$.
Both the background flow velocity at the level of the vortices and the phase velocity of the linear instability
are also approximately equal to $1$.
Animations  of both evolving flows are available as supplementary materials.

\section{Discussion and Conclusions}
\label{sec:conclusion}

In this paper, we have described a new, inherently  viscous instability and have demonstrated that it shares many of the characteristic
features of the classic, inviscid Holmboe wave instability, namely manifesting as a propagating vortex on either side of the mixing layer
and appearing to be caused by the interaction of internal gravity waves on a shear interface.
Since it exists in regions of parameter space where no instability is predicted in the inviscid limit,
we term it the viscous Holmboe instability, or VHI.
The instability we have described is distinct from the `viscous Holmboe wave instability' found by \citet{eaves_caulfield_2017} in  plane Couette flow,
which required non-slip and non-penetration effects in the presence of a rigid boundary, whereas we have shown that boundaries only weakly affect the instability,
and the VHI discussed here is truly an instability of a stratified shear layer.
Despite the similarities to inviscid HWI, it has significant differences from the classical case:
it exists when the density interface is \emph{not sharp} compared with the shear layer;
it can have a phase speed greater than the maximum fluid velocity;
and it is \emph{destabilised} by viscosity.
As a result of the fact that the VHI described here is an instability
of an unsteady background profile which itself diffuses away slowly, we find the curious situation
that although this is an instability which requires viscosity to exist, the effect of the instability
relative to the diffusion of the background flow appears to be greater as $Re$ is increased.

This work is a study of how viscosity affects the Holmboe wave instability as certain parameters
are varied. There are many possible extensions which have been examined for (classic) HWI,
including considering the effects of compressibility \citep{witzke2015shear} and relaxing the Boussinesq
approximation \citep{umurhan2007holmboe, churilov2019}.
We briefly investigated the possibility that the higher Holmboe modes described by
\citet{alexakis2005holmboe,alexakis2007,alexakis2009stratified} are also destabilised by viscosity at low $R$,
and did indeed find a further band of instability with very small growth rates.
Our work has been entirely restricted to two dimensions. Though this is a common
assumption when studying linear instabilities of shear flows, there is no physical basis for this, and indeed we would fully expect to see the fastest growing
mode being three-dimensional  in some regions of parameter space, based on the results of \citet{smyth_peltier_1990}.
A third dimension would also significantly affect the nonlinear evolution of the instability at high $Re$.

Despite the lack of a sharp density interface relative to the shear layer for the parameters for which
we have found instability, we would certainly still expect internal gravity waves to be present
on the interface. There is no reason we are aware of, \textit{a priori}, to think that these could not resonate
with the vorticity waves to cause instability. The wave resonance descriptions of stratified shear
instabilities have been mainly qualitative, except in the cases of piecewise constant density and vorticity profiles,
which would be physically inconsistent at finite viscosity.
Recent attempts to analyse the components of resonances \citep{carpenter_2010, eaves_balmforth_2019}
and to understand better the dynamics of the resonant system \citep{heifetz_guha_2018,heifetz_guha_2019}
have relied on analysis which requires perturbations to be inviscid, and these certainly would not
apply in the low $Re$ regimes we have described.
Though the theory of wave resonance has given useful insight in many situations, it is clearly not the full picture.
One major outstanding question is how the Miles-Howard criterion may relate to the wave resonance picture.
\citet{baines1994mechanism} give an argument from critical layer theory, though the authors themselves
admit that this gives neither a necessary nor sufficient criterion for stability.

Most of the unstable regions of the viscous Holmboe instability for $R\leq 2$ have $|c_r|>1$,
so there is no critical layer. Therefore, Lindzen's wave over-reflection hypothesis for the mechanism
of stratified shear instabilities, as well as other interpretations based on the existence of a critical
layer, such as the wave-particle interaction described by \citet{churilov2019}, cannot apply.
This is in contrast with the viscous instability described by \citet{miller_lindzen_1988}, in which
the viscosity was thought to enable over-reflection at the critical layer.
As discussed by \citet{smyth1989transition}, it could be possible that the instability
is associated with over-reflection of a wave with a different phase speed, which therefore could itself have
a critical layer, but this makes an intuitive explanation much harder.
Since the wave over-reflection theory is not a predictive explanation of the instability in this case,
it does not seem useful here, though it has certainly proven important in many other circumstances.

Under carefully controlled parameters, we have been able to show significant nonlinear growth of the viscous
Holmboe instability at $R=1.5$ and $Re=4000$, from initial noise, leading to secondary instabilities and transition to disorder.
This primary instability has no critical layer.
Nevertheless, most of the regions of instability we have studied, with $R<2$, have much lower growth rates.
We conclude that the viscous Holmboe instability is unlikely to be particularly significant
in physical processes. In addition to this, for typical values of Prandtl number in the atmosphere ($Pr\approx0.7$)
we see very small growth rates and for typical values of  $Pr \approx 7$ in the oceans,
we see the full classical HWI, since in this case $R$ is usually large.

Despite these caveats, we have demonstrated the definite existence of an instability which bears a striking
resemblance to HWI, but violates many of the supposed prerequisite conditions.
We therefore suggest that any instability in a stratified shear layer be considered
Holmboe instability if it manifests as propagating vortices on either side of the shear layer,
regardless of the relative width of the density interface, the presence of critical layers or the
minimum value of the gradient Richardson number.

\section{Declaration of Interests}
The authors report no conflict of interest.

\bibliographystyle{jfm}
\bibliography{pck20}

\end{document}